# Uncovering cooling center usage as an adaptation strategy for hurricane-blackout-heat compound hazards during Hurricane Beryl (2024)

Tianle Duan[1, *], Fengxiu Zhang[2], Qingchun Li[1]

[1]School of Construction Management Technology, Purdue University, West Lafayette, IN, USA

[2]Schar School of Policy and Government, George Mason University, Arlington, VA, USA

## Abstract

Extreme heat and hurricane-induced blackouts could occur simultaneously in the summer, posing great challenges to community health and well-being. Cooling centers serve as a key adaptation strategy to alleviate heat stress, especially among heat-vulnerable populations. This study leverages mobility data to examine how affected communities utilize cooling centers in response to hurricane-blackout-heat compound hazards. Additionally, it examines disparities in cooling center usage, focusing on individuals with access and functional needs (AFNs) who are usually overlooked in emergency management practices. These populations include but are limited to older adults, individuals with limited English proficiency, people with disabilities, those without vehicle access, and lower-income households. Using the empirical case of Hurricane Beryl (2024) in Harris County, Texas, we find no statistically significant difference in visiting formal (established by the government) and informal cooling centers (operated by volunteer organizations). Census block groups closer to the nearest cooling center and those with lower income are more likely to seek shelter from extreme heat at cooling centers in the aftermath of Hurricane Beryl. Lower-income block groups also tend to be situated closer to cooling centers, suggesting that Harris County may have strategically placed them in areas with greater social vulnerability. Furthermore, we investigate visiting hotels as an alternative but more expensive adaptation strategy during Hurricane Beryl. Between these two adaptation options, shorter distances to cooling centers, lower income, and elder age are statistically significantly associated with a higher probability of visiting cooling centers rather than hotels, while limited English proficiency significantly decreases such probability. These findings underscore the importance of equitable access to cooling centers and highlight the need for targeted outreach and support to ensure that people with AFNs can effectively utilize these critical adaptation resources.

1. **Introduction**

Compound hazards refer to the combination of two or more hazardous events that occur simultaneously or sequentially, often interacting to amplify their individual impacts[1]. Ongoing climate change has heightened the complexity and frequency of compound hazards, introducing new combinations of weather extremes and the impacts they impose[2]. A particularly concerning and increasingly recognized class of compound hazards involves major power outages and extreme heat following major hurricanes. Massive power outages have already occurred in the aftermath of several hurricanes, such as 2012's Hurricane Sandy and 2017's Hurricanes Irma and Maria, incurring power outages to between 3.3 million and 8.2 million customers, respectively[3,4]. Most recently, Hurricane Beryl caused over 2.7 million Houstonians to lose power, with more than 1.5 million still without power 24 hours after the hurricane had left Houston[5]. When coinciding with extreme heat, hurricane-induced outages significantly burden affected communities by critically limiting access to air conditioning and other cooling systems essential for coping with heat stress. Population exposure to extreme heat both inside and within buildings can reach dangerous levels as mechanical air conditioning systems become inoperable[6].

Extreme heat is already the deadliest weather-related hazard in the United States, causing more fatalities annually than any other type of extreme weather[7,8]. Prolonged exposure to extreme heat can lead to heat-related illnesses such as heatstroke, dehydration, and cardiovascular stress, particularly among vulnerable populations like the elderly, children, the homeless, and those with pre-existing health conditions[9,10]. A recent study estimates that extreme heat is responsible for approximately 12,000 premature deaths per year in the current decade[8]. Continued warming due to climate change is expected to significantly increase heat-related morbidity and mortality by mid-century[8,11], the risks of which are further amplified if we account for hurricane-induced grid failures that coincide with extreme heat[12]. A recent study finds that simulated compound events of grid failure and extreme heat can expose 68% to 100% of the urban population to an elevated risk of heat exhaustion and/or heat stroke[6].

Cooling centers have been employed as a key strategy for mitigating the health impacts of extreme heat, especially among those without access to air conditioning systems or those who cannot afford them[9,13]. Cooling centers are publicly accessible, climate-controlled spaces where large groups of people can seek shelter from extreme heat. These centers may be formally designated or informally operated by volunteering organizations. Formal cooling centers are typically operated by government agencies and are often located in community centers, schools, or municipal buildings to serve as designated heat refuges. Informal cooling centers include spaces such as shopping malls, faith-based organizations, humanitarian facilities, and movie theaters. By offering temporary relief, cooling centers help to prevent heat-related illnesses such as dehydration, heat exhaustion, and heat stroke. Compound hazards involving hurricane-induced power outages and extreme heat further heighten the importance of cooling centers. In such scenarios, they provide essential protection for heat-vulnerable populations, including older adults, low-income individuals, people experiencing homelessness, those without vehicles, and individuals with pre-existing health conditions[9,10]. In addition, they can function as emergency shelters, providing access to critical health information, supportive services, and essential resources such as drinking water, food, and electricity for charging devices[9,14].

The critical role of cooling centers in heat mitigation has prompted extensive research on their accessibility and the barriers that prevent people from accessing them. Studies have examined the geographic distribution, proximity, and availability of cooling centers, revealing significant disparities in access across different populations. For instance, research in New York City revealed that only one-third of the general population and 50% of heat-vulnerable neighborhoods were located within walking distance (0.5 miles) of a cooling center[15]. Similarly, a recent study on cooling centers in the Southeastern U.S. states found that only 36 percent of the general population lives within a 15-minute drive of a cooling center and, in most states, less than 10 percent of vulnerable populations are within this drive shed[16]. Additionally, various physical, operational, social, and cultural barriers hinder individuals from effectively accessing and utilizing cooling centers. Physical barriers include inadequate transportation options, long distances to centers, and limited infrastructure in underserved areas[17,18]. Long distances are particularly challenging for individuals with mobility limitations and underlying health conditions. Operational barriers involve factors such as restricted hours, insufficient staffing, admission costs, or overcrowding during extreme heat events[9,10]. Social and cultural barriers can include language challenges, a lack of awareness about available cooling centers, stigma or distrust towards government-run services, and cultural differences that prevent certain groups from seeking help[10,17].

Location-based data offers distinct advantages in real-time responses and fine spatiotemporal resolution, making it well-suited for studying the population's adaptation behavior under compound hazards. Existing studies have frequently used location-based data (big mobility data) to investigate how people prepare for, respond to, and recover from adverse events. Dargin et al. leveraged GPS trajectory data to investigate people's footprints during Hurricane Harvey, finding a statistically significant increase in visits to specific points of interest (e.g., gasoline stations, grocery stores, and insurance carriers) before Harvey[19]. Li and Mostafavi[20] used mobility data to reveal the extent, timing, and spatial variation of preparedness patterns for Hurricane Harvey in Harris County, Texas. Li et al. used mobility data to investigate income and race inequality in preparedness for Hurricane Ida (2021) in Louisiana[21]. Bian et al. used gasoline station visit data during Hurricane Ida (2021) to explore factors affecting the fuel demand deviation at an aggregate level[22]. With specific relevance to extreme heat and cooling centers, Tian et al. used mobile phone location data to examine how individuals in diverse communities adjust their travel behavior in response to extreme heat, identifying distinct patterns during both daytime and evening hours[23]. Similarly, Derakhshan et al. employed smartphone location data to reveal visit and utilization patterns of cooling center, while also exploring how these patterns vary across communities with different social vulnerability indices[18].

Nevertheless, in practice, we have limited knowledge of how cooling centers are accessed during compound hazards—like hurricanes accompanied by power outages and extreme heat—as a heat mitigation strategy. Critical questions persist regarding cooling centers, usage patterns, the sociodemographic profiles of individuals visiting these centers, as well as the factors that influence people's adaptation preferences. Of particular concern is cooling center accessibility and visits for vulnerable populations. Specifically, this study will pay special attention to individuals with access and functional needs (AFNs), defined as individuals who require additional support due to physical, cognitive, or health-related conditions. Those include people with disabilities, chronic illnesses, limited mobility, or age-related factors that have often been neglected in disaster management planning and operations[24,25].

To address these gaps, this research will leverage location-based data to analyze the mobility patterns of individuals during periods of simultaneous outages and extreme heat, using Hurricane Beryl and Harris County, TX as a case study. As such, this study contributes a nuanced understanding of accessibility and visit patterns of cooling centers when people simultaneously confront hurricane-induced power outages and extreme heat. It sheds light on the effectiveness of cooling centers in reaching those most in need and identifies factors that hinder equitable access during such compound crises. By identifying potential barriers faced by groups with diverse AFNs, this study aims to inform the development of more inclusive and equitable management strategies for cooling centers and other emergency resources. In Section 2, we introduce the data and methods used in this study. Section 3 presents the results, followed by a discussion of our findings and limitations in Section 4.

## 2. Data and methods
### 2.1 Foot traffic data

The foot traffic data for our analysis is provided by Dewey Inc. by subscription. This dataset included daily visit counts for each point of interest (POI) along with additional attributes such as geometric polygons, geographic coordinates, categories, and North American Industry Classification System (NAICS) codes. POIs are physical establishments that someone may find helpful or interesting, such as restaurants, gasoline stations, and libraries. The foot traffic data computes visits, visitors, and other metrics within a POI's geometry based on the location data from visitors' devices. By grouping the POIs by category, we analyzed changes in visits to cooling centers during the Hurricane Beryl event to investigate people's adaptation to the compounded hazards.

**2.2 Location-based data**

We matched cooling centers to the POIs in the Dewey foot traffic data based on the following processing. We compiled the list of cooling centers from four sources: the City of Houston Office of Emergency Management, Houston Public Media, the Houston Chronicle, and KHOU, and then removed the duplicates. The list includes details of cooling centers, such as names, categories, street addresses, and operating hours. We used the Nominatim geocoding service [26] and Google Maps to extract latitude and longitude coordinates from the provided street addresses of cooling centers and cross-validated each other. Then, we matched the coordinates of cooling centers to the POIs in the foot traffic dataset. If the geographic coordinates of a cooling center fell within a POI polygon, we linked the cooling center to the corresponding POI. Otherwise, we identified the five closest POIs to the cooling center based on geographic proximity and manually verified the matched POI by cross-referencing the POI category. We filtered Hotels' location-based data from the Dewey foot traffic dataset using NAICS code 721100.

We calculated the Euclidean distance between each census block group (CBG) and its nearest cooling center based on the coordinates of the matched POIs and CBG centroids. We used the Euclidean distance between a CBG and a cooling center as a critical confounding variable affecting cooling center usage.

**2.3 American Community Survey Data**

To study the factors influencing adaptation preference, we considered socioeconomic variables and AFNs. Individuals with AFNs require additional assistance before, during, and after emergencies due to limitations in mobility, communication, transportation, medical care, or other essential services. These needs may arise from disabilities, chronic health conditions, age-related factors, or other circumstances that affect their ability to respond to and recover from disasters[24]. As emergency management practices have primarily viewed vulnerability through the lens of wealth and race[27], people with AFNs have often been overlooked in planning and response efforts[25,28].

To fill the gap in research and practice, we considered the following AFN variables including household median income, percentage of households with limited English Proficiency, percentage of households without vehicles, percentage of people over 65, percentage of people under 5 as well as percentage of households with individuals living with disabilities. Previous studies consistently demonstrated that higher median income is correlated with more proactive behavior toward hazardous events[29,30]. These groups often possess higher adaptive capacity due to their access to critical resources and networks which enable more effective preparation, response, and recovery during hazardous events. English proficiency is an important factor in assessing the ability to understand advisories from the National Hurricane Center (NHC) and local governments[31], while vehicle availability significantly influences mobility during emergencies. All data were obtained from the American Community Survey provided by the U.S. Census Bureau.

## 2.4 Statistical Analysis

In the aftermath of Beryl, during a massive power outage and extreme heat, short of relocating to another area, individuals can adapt by seeking refuge in a hotel or a cooling center. We conducted statistical analyses to explore the AFN factors influencing people's adaptation strategies. We employed the Kruskal-Wallis test and Dunn's test to examine the distribution of cooling centers across different income groups. We conducted the Mann-Whitney U test (MW test) to examine the differences in AFN factors across CBGs exhibiting two adaptation strategies. Further, we conducted logistic regression analysis to investigate the roles of AFN factors in adopting adaptation strategies by controlling for potential interactions among the variables.

## 3. Results
### 3.1 Hurricane Beryl (2024) and Harris County, TX

Hurricane Beryl (2024) made landfall near Matagorda, Texas, about 85 miles south-southwest of Houston, on July 8, 2024, and extended to Harris County in a few hours[32]. Beryl exemplified a compound hazard, where multiple hazards converge to amplify damage and losses in affected communities. It unleashed sustained destructive winds up to 80 miles per hour and triggered significant coastal and flash flooding, causing extensive damage across the region[32]. In the meantime, more than 2.7 million customers experienced widespread power outages[33]. Roughly 87% of the electricity customers in Harris County lost power after Hurricane Beryl tore through early 8$^{th}$ July[5]. In the aftermath, extreme heat— with temperatures reaching the mid-90s (°F) and a heat index exceeding 100°F—further intensified the crisis in Harris County, compounding the hazards and serious threats to public health and safety[34].

In the following section, we investigate the distribution of cooling centers, visits to both formal and informal cooling centers, preferences between cooling centers and hotels for adaptation, as well as the factors influencing people's adaptation strategies.

### 3.2 Cooling center distribution in Harris County

We matched 101 cooling centers in Harris County in the foot traffic data (see Section 2.2 for method details). **Fig. 1** illustrates the distribution of cooling centers in Harris County. Among them, 54 are formal cooling centers (i.e., established by governments) and 47 are informal cooling centers (i.e., operated by nonprofit or commercial organizations). We presented the list of cooling centers in the Supplementary Material.

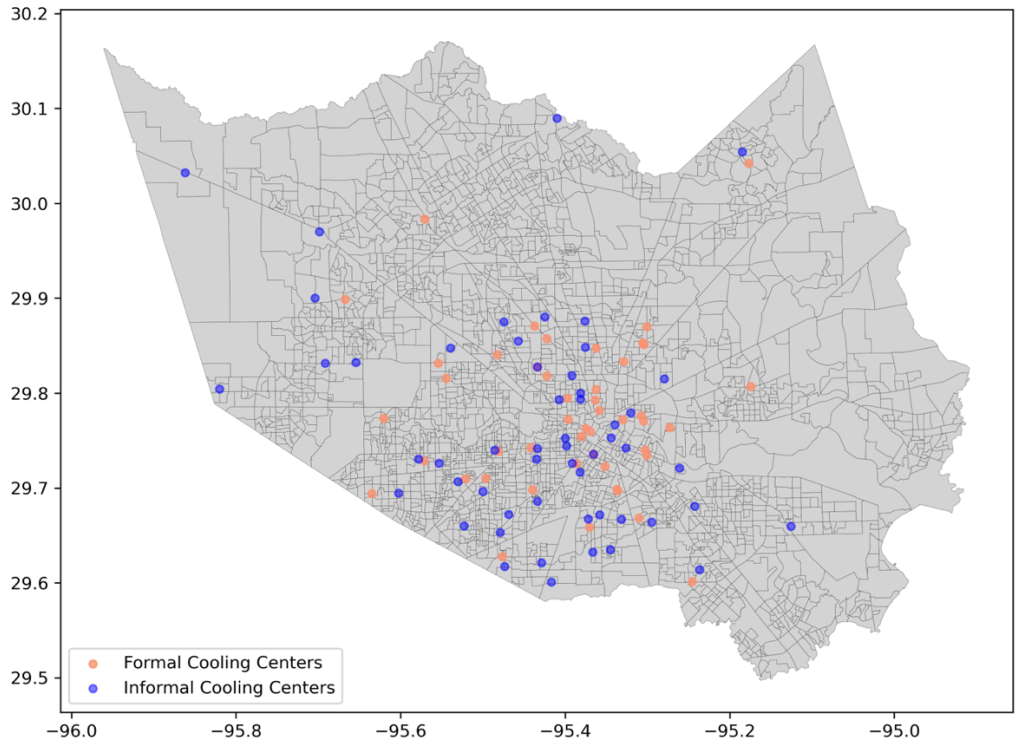

**Fig. 1 Cooling centers in Harris County, TX**. 54 formal cooling centers and 47 informal cooling centers.

To examine how cooling center distribution differs across CBGs, we calculated the Spearman correlation coefficients among AFN factors and presented the results in Fig. 2. The percentage of households with limited English proficiency and the percentage of households without a vehicle all show strong correlations with household median income, with Spearman correlation coefficients of -0.51 (p <0.001) and -0.49 (p<0.001), respectively. Therefore, we used household median income as a representative variable to further assess the equity of cooling center distribution.

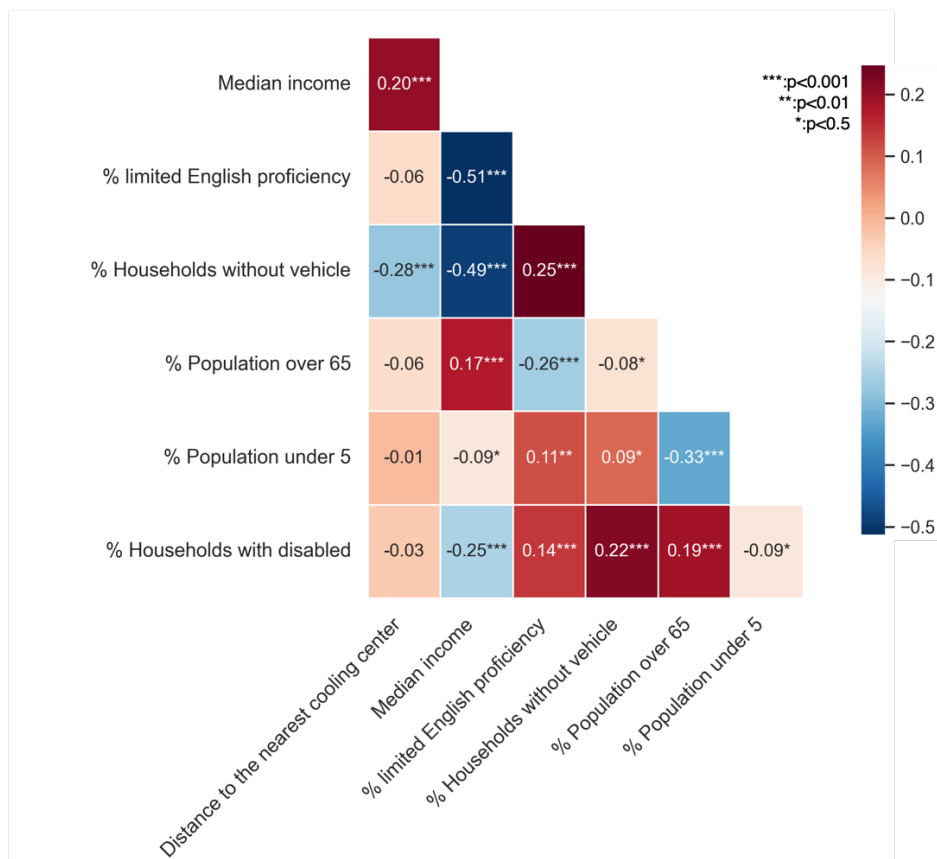

**Fig. 2 Spearman correlation heatmap**

We then divided the 2830 CBGs in Harris County into three groups based on household median income: low (N=944), medium (N=943), and high-income groups (N=943), with the high-income group representing the top 33.3% of CBGs. We performed a Kruskal-Wallis test to evaluate differences in the distances to cooling centers across the three CBG groups, yielding a p-value < 0.001, indicating statistically significant differences. To identify which groups differed significantly, we conducted post-hoc Dunn's tests between each pair of two CBG groups. **Fig. 3** illustrates statistically significant differences between each pair of groups ($p < 0.05$ between the low-medium pair and medium-high pair, and $p < 0.001$ between the low-high pair). The high-income group exhibited the longest distance to the nearest cooling centers, followed by the medium-income group, with the low-income group having the shortest distance.

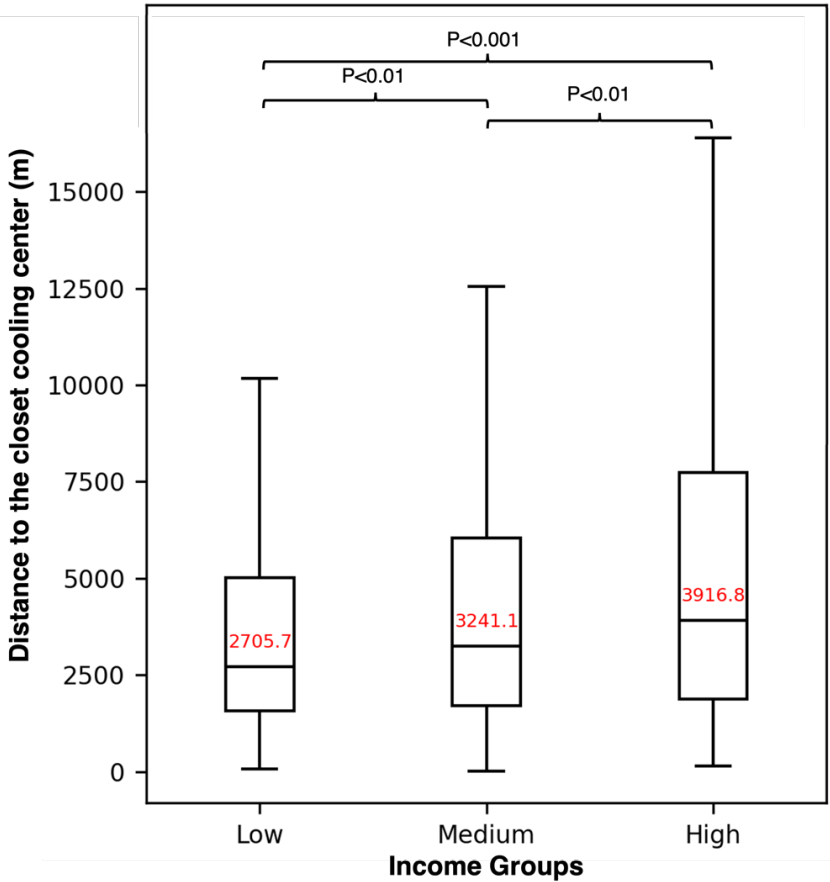

**Fig. 3 Distance to the closest cooling center across income groups.** Red numbers in the figure represent the median value of each income group.

### 3.3 Formal and informal cooling center visits

To analyze people's visits to cooling centers after Hurricane Beryl, we computed and compared the weekly visits to cooling centers before Beryl's landfall (1 July to 7 July) and after (8 July to 15 July). Previous studies have demonstrated a decreasing tendency for POI visits during and shortly after hazardous events, as people reduce unnecessary outings and transportation disruptions limit access to destinations[21,35]. Therefore, an increase in visits to cooling centers during power outages suggests adaptive behavior by the population to mitigate the impacts on their health. We divided the CBGs in Harris County into two groups based on their residents' tendency (i.e., a greater increase in visits) to visit formal cooling centers and informal cooling centers. We conducted a Mann-Whitney U test between the two groups to examine how visits to the formal and informal cooling center vary across the six AFN categories and presented results in Fig. 4. Results show that there is no statistically significant difference in the visits of different types of cooling centers across these AFN categories at the 95% confidence level. This suggests that people accessed informal cooling centers at similar rates to formal cooling centers to mitigate heat stress. In the following analysis, we combined the formal and informal cooling centers when compared with alternative adaptation choices (i.e., hotels).

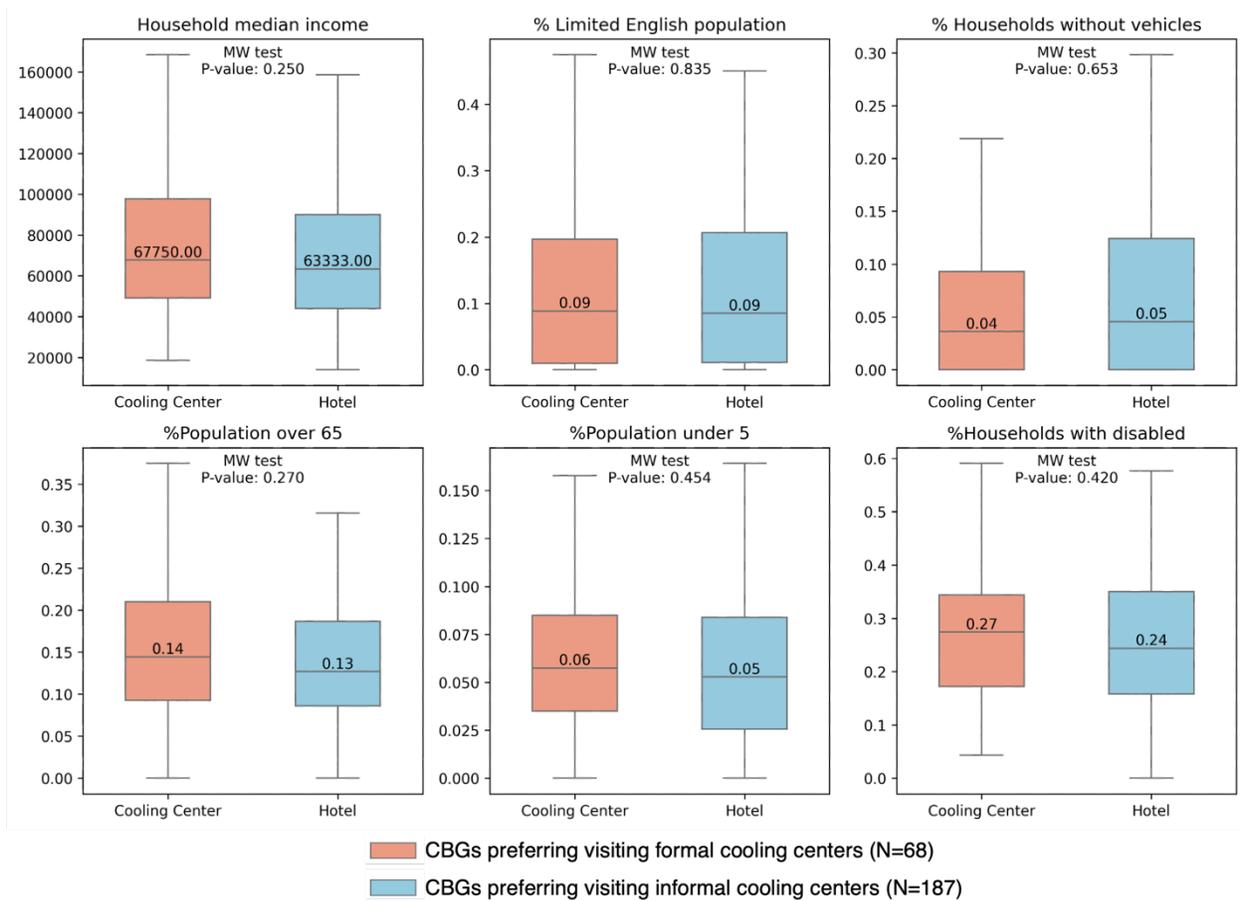

**Fig. 4 Mann-Whitney U Test comparing formal and informal cooling center visits across AFN categories**

### 3.4 Cooling center and hotel visits

Besides shelter in cooling centers for adaptation, shelter in hotels is an alternative adaptation strategy during power outages and other infrastructure system disruptions[36–38]. After Beryl, 170 CBGs in Harris County exhibited adaptive behaviors by visiting cooling centers (highlighted in blue in Fig. 5a). As an alternative, 614 CBGs showed adaptive behavior by visiting hotels (highlighted in red in Fig. 5b). 134 CBGs showed increasing visits to both cooling centers and hotels. By comparing the ratios of increased visits to cooling centers and hotels, we determined that a higher ratio indicates a stronger adaptive preference for the corresponding choice. We visualized the CBG-level preferences for adaptive behaviors (135 to cooling centers versus 635 to hotels) in Fig. 5c.

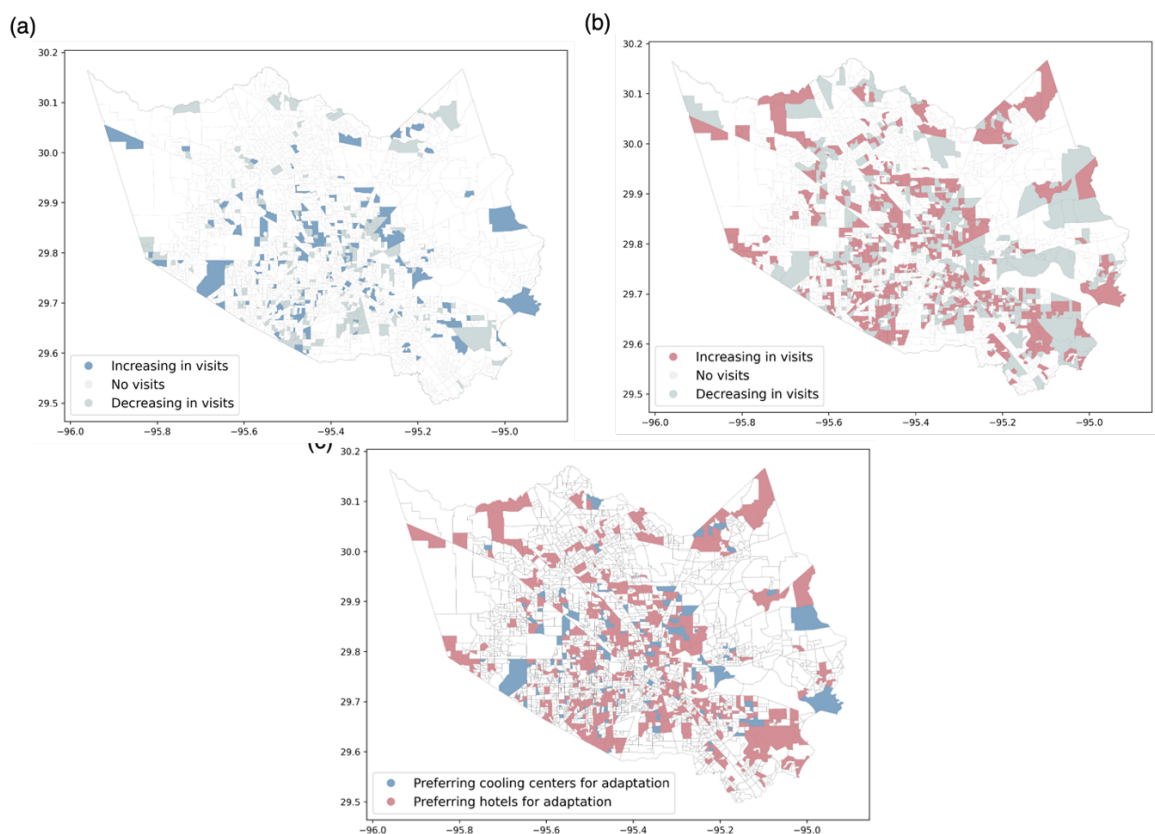

**Fig. 5 POI visit trends**: (a) Changes in cooling center visits (Increasing (N=170) Vs. Decreasing (N=208)); (b) Changes in hotel visits (Increasing (N=614 Vs. Decreasing (N=429)); (c) Adaptation preference (Cooling centers (N=135) Vs. Hotels (N=635), 14 CBGs showing equal preference)

We conducted Mann-Whitney U tests to examine the differences in distance between the cooling centers between CBGs with distinct adaptation preferences. In **Fig. 6**, boxplots illustrate CBGs with a preference for cooling centers in orange and those preferring hotels in blue. The Mann-Whitney U test for differences in distances to the nearest cooling center yielded a p-value of 0.029, indicating statistically significant differences between the two clusters —CBGs with a preference for cooling centers have a closer distance to its closest cooling centers. Fig. 7 presents the results of the Mann-Whitney U tests examining AFN factors between the two clusters. We observed statistically significant differences in CBG-level household median income — CBGs that prefer cooling centers tend to have lower median incomes ($64,327 versus $67,750). Moreover, CBGs with a higher proportion of households without vehicles (5% vs. 3%) show a stronger preference for cooling centers compared to those in the opposite cluster. There is no evidence for statistical differences in English proficiency, age structure, and the percentage of households with people with disabilities. The results indicate that individuals with lower socioeconomic status or with no vehicles are more likely to prefer cooling centers as an adaptation option during power outages, whereas those with higher socioeconomic status tend to visit hotels.

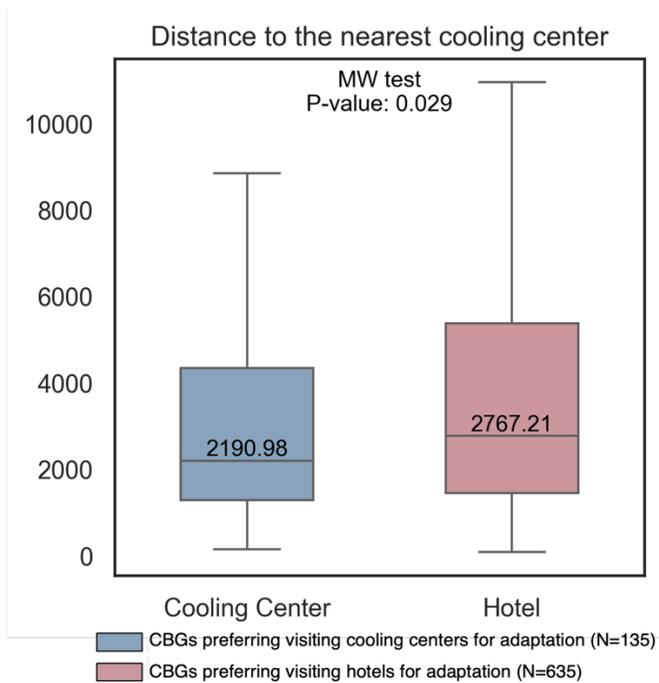

**Fig. 6 Mann-Whitney U test for differences in distances to the nearest cooling center across two preference CBG groups**

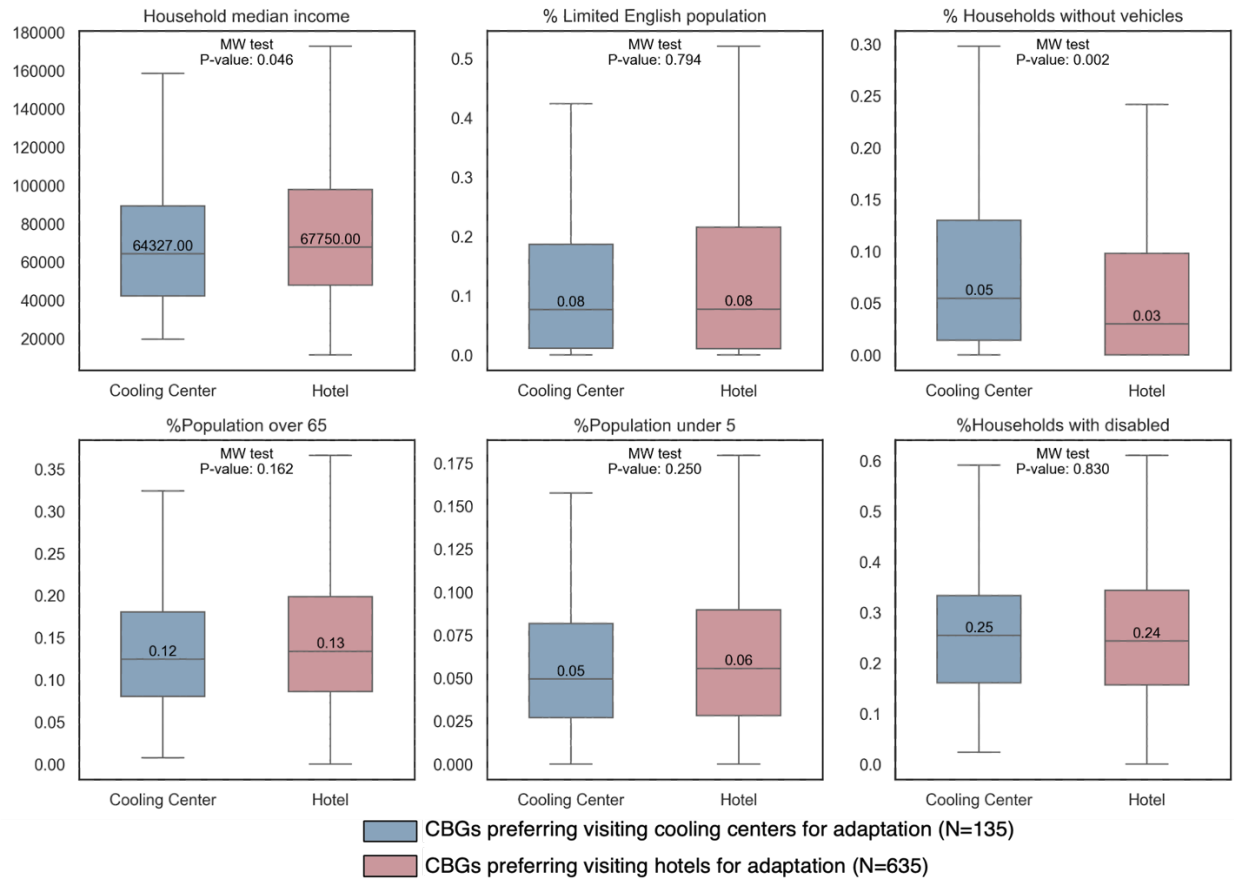

**Fig. 7 Mann-Whitney U Test Comparing AFN factors across two preference CBG groups**

## 3.5 AFN factors in adaptation strategies

To examine the relationships between various AFN factors and the adoption of adaptation strategies, we performed two logistic regression models with different dependent variables. Model 1 in Table 1 predicts visits to cooling centers among all CBGs (N=2830), coding CBGs using cooling centers for adaptation as 1 and those without using cooling centers as 0. Results show that the distance to the closest cooling center and the household median income exhibits a significantly negative correlation with the probability of visiting cooling centers for adaptation, suggesting that geographical proximity facilitates the use of cooling centers, and the lower income groups tend to use cooling centers.

Model 2 examines factors influencing CBGs that visited cooling centers (N=770), specifically exploring why they prefer cooling centers over hotels. We coded CBGs preferring visiting cooling centers as adaptation strategy as 1 and those preferring hotels as 0. Results show that CBGs with shorter distances to cooling centers, lower household median income, a lower percentage of households with limited English Proficiency, and a higher percentage of the population over 65 are more likely to prefer visiting cooling centers as an adaptation strategy. The findings in the two logistic models consistently suggest that people with AFNs tend to use cooling centers to adapt to the blackout-heat compound hazards.

**Table 1 Logistic regression results for adaptation strategies**

|  | Model1 | Model2 |
|---|---|---|
|  | Visiting cooling centers for adaptation | Prefer cooling centers to hotels |
| Distance to the closest cooling center (m) | -0.372(0.086)*** | -0.167(0.098)* |
| Household median income (in $1,000) | -0.250(0.126)** | -0.216(0.129)* |
| % Households with limited English Proficiency | -0.124(0.102) | -0.206(0.112)* |
| % Households without vehicles | 0.040(0.090) | 0.134(0.102) |
| % Population over 65 years | 0.028(0.095) | -0.198(0.111)* |
| % Population under 5 years | -0.092(0.096) | -0.166(0.104) |
| % Households with disabled members | -0.120(0.093) | -0.033(0.102) |
| N | 2830 | 770 |
| Pseudo R-squared | 0.028 | 0.028 |

***p-value<0.01, **p-value<0.05, *p-value<0.1

## 4. Discussion and conclusion

Compound hazards involving power outages and extreme heat in the aftermath of hurricanes are becoming increasingly frequent and severe. Given the vital role that cooling centers play in mitigating heat stress—especially for vulnerable populations—this study examines the distribution and visits of cooling centers in the context of hurricane-induced power outages and extreme heat. We further investigate the characteristics of social groups visiting cooling centers under such scenarios, with a particular focus on individuals with AFNs. The results highlight key patterns, revealing that while cooling centers serve as an essential heat mitigation strategy, their accessibility and usage remain uneven across different demographic groups.

Using the case study of Hurricane Beryl, our study unravels key patterns of cooling center distribution. Informal cooling centers are more widely distributed and are accessible to a larger fraction of CBGs than formal cooling centers in our study area. Furthermore, CBGs with lower income are shown to be more proximate to the closest cooling centers. This finding contrasts with previous findings that suggest lower accessibility among heat-vulnerable neighborhoods [16], indicating that cooling centers in Harris County may be strategically placed in areas with greater social vulnerability. However, we were unable to determine whether the cooling centers were established as permanent facilities or activated temporarily in response to Hurricane Beryl's aftermath to evaluate their long-term availability and reliability.

Regarding the usage of cooling centers, our results show no significant difference in CBGs' visits to formal and informal cooling centers, implying that both types play an equally important role in providing relief during extreme heat events. This finding highlights the critical function of informal cooling centers in supplementing formal facilities, potentially filling gaps in accessibility and capacity. Their importance becomes even more pronounced when considering their wider distribution and easier access, as demonstrated in our analysis.

When combining the distribution and usage of cooling centers, our results show that CBGs closer to cooling centers tend to visit cooling centers as heat refugee during extreme heat. This result reinforces previous findings on the crucial role of accessibility in promoting usage and enhancing community resilience to extreme heat[15,17,18]. Additionally, CBGs with lower median household incomes are also more inclined to visit cooling centers, aligning with recent studies indicating that vulnerable communities make more frequent use of cooling centers to mitigate heat impacts[18]. Economic factors influence the choice of cooling center use, possibly due to the affordability and availability of alternative cooling options. These findings highlight the intersection of geographical and socioeconomic factors in shaping the adaptive behaviors of communities facing extreme heat.

Furthermore, our breakdown analysis of cooling center visit preferences across AFN groups reveals that CBGs with lower income tend to prefer cooling centers over hotels as their primary adaptation strategy. This preference may be driven by financial constraints and greater reliance on publicly available resources. Additionally, our examination of the effects of AFN factors highlights the various barriers faced by individuals in several AFN categories. CBGs with a higher proportion of individuals with limited English proficiency, older adults, and children overall are less likely to prefer cooling centers as an adaptation strategy. Communication challenges, mobility limitations, and safety concerns could all hinder their usage of cooling centers. Moreover, although our results show a relatively equitable distribution of cooling centers, the breakdown analysis demonstrates that some social groups, especially those in some AFN categories, still experienced compound barriers in accessing those facilities. As a consequence, those individuals may face elevated risks of heat-related illnesses and other adverse health outcomes due to their constrained ability to seek heat refuge. This disparity underscores the need for targeted interventions, including enhanced transportation options, multilingual outreach efforts, and accessibility improvements, to improve the effectiveness of cooling centers in serving all vulnerable populations.

Our study highlights several key avenues for future research. To begin with, while our mobility data reveals whether each individual visited a cooling center, we had no insight into the duration of each visit. We also aggregated visits across the study timeframe, without distinguishing the frequency or timing of visits within specific heat periods or operating hours of the cooling centers. Future research could explore these aspects by tracking visit duration, frequency, and timing to gain a more nuanced understanding of how cooling centers are utilized during different conditions and across diverse populations. Moreover, while we inferred individuals' preferences for cooling centers during heat events based on accessibility and usage patterns, future studies should focus on developing more direct measures of individual decision-making processes, motivations, and preferences. Resident surveys and interviews could offer valuable insights into the factors influencing cooling center preferences and selection, with a particular focus on the challenges faced by individuals with AFNs. Additionally, our analysis is limited by the availability and granularity of data, which may not sufficiently capture community-specific or individual-specific variations. Relatedly, our snapshot data, which was collected in the aftermath of a hurricane, may not accurately reflect the availability and usage of cooling centers under routine conditions. The circumstances following a disaster, such as surged demand, increased supply, or temporary changes in infrastructure, could influence the availability and usage of cooling centers, making it difficult to generalize these findings to non-crisis situations. Future research should consider longitudinal studies that assess both routine and emergency contexts to better understand how cooling center availability and usage vary across different conditions and communities.

## 5. Author contributions

Q.L., F.Z.: Conceptualization; T.L., F.Z.: Methodology; T.L.: Analysis, visualization; T.L., F.Z. writing – Original draft; T.L., F.Z., Q.L..: Writing – Reviewing and Editing

## 6. Competing interest

The authors declare no competing interest.

# 7. Acknowledgement

The authors would like to acknowledge funding support from the Public Health Extreme Events Research (PHEER): "Uncovering Compound Hazards Adaptation during Hurricane Beryl Using POI Visit Data."

10. Sampson, N. R. *et al.* Staying cool in a changing climate: Reaching vulnerable populations during heat events. *Glob Environ Change* **23**, 475–484 (2013).

11. Bobb, J. F., Peng, R. D., Bell, M. L. & Dominici, F. Heat-Related Mortality and Adaptation to Heat in the United States. *Environmental Health Perspectives* **122**, 811–816 (2014).

12. Matthews, T., Wilby, R. L. & Murphy, C. An emerging tropical cyclone–deadly heat compound hazard. *Nature Climate Change* **9**, 602–606 (2019).

13. Uejio, C. K. *et al.* Intra-urban societal vulnerability to extreme heat: the role of heat exposure and the built environment, socioeconomics, and neighborhood stability. *Health & place* **17**, 498–507 (2011).

14. Fraser, A. M. *et al.* Household accessibility to heat refuges: Residential air conditioning, public cooled space, and walkability. *Environment and Planning B: Urban Analytics and City Science* **44**, 1036–1055 (2017).

15. Nayak, S. G. *et al.* Development of a heat vulnerability index for New York State. *Public Health* **161**, 127–137 (2018).

16. Allen, M. J., Whytlaw, J. L., Hutton, N. & Hoffman, J. S. Heat Mitigation in the Southeastern United States: Are Cooling Centers Equitable and Strategic? *Southeastern Geographer* **63**, 366–385 (2023).

17. Bedi, N. S., Adams, Q. H., Hess, J. J. & Wellenius, G. A. The Role of Cooling Centers in Protecting Vulnerable Individuals from Extreme Heat. *Epidemiology* **33**, 611 (2022).

18. Derakhshan, S. *et al.* Smartphone locations reveal patterns of cooling center use as a heat mitigation strategy. *Applied Geography* **150**, 102821 (2023).

19. Dargin, J. & Mostafavi, A. Well-Being and Infrastructure Disruptions during Disasters: An Empirical Analysis of Household Impact Disparities during Hurricane Harvey. in

28. Peek, L. & Stough, L. M. Children With Disabilities in the Context of Disaster: A Social Vulnerability Perspective. *Child Development* **81**, 1260–1270 (2010).

29. Cutter, S. L., Boruff, B. J. & Shirley, W. L. Social vulnerability to environmental hazards. *Social science quarterly* **84**, 242–261 (2003).

30. Miao, Q. & Zhang, F. Drivers of Household Preparedness for Natural Hazards: The Mediating Role of Perceived Coping Efficacy. *Natural Hazards Review* **24**, 04023010 (2023).

31. Xiang, T., Gerber, B. J. & Zhang, F. Language access in emergency and disaster preparedness: An assessment of local government "whole community" efforts in the United States. *International Journal of Disaster Risk Reduction* **55**, 102072 (2021).

32. National Weather Service. Hurricane Beryl 2024. *National Weather Service, National Oceanic and Atmospheric Administration* https://www.weather.gov/lch/2024Beryl (2024).

33. Wolfe, S. More than 2.5 million without power in Texas as tropical storm Beryl pushes inland. *POWERGRID International* https://www.power-grid.com/td/outage-management/more-than-2-5-million-without-power-in-texas-as-tropical-storm-beryl-pushes-inland/ (2024).

34. Harvey, C. How Heat Combined with Hurricane Beryl to Cause Misery in Houston. *Scientific American* https://www.scientificamerican.com/article/heat-combined-with-hurricane-beryl-to-cause-misery-in-houston/ (2024).

35. Zhang, X. & Li, N. Characterizing individual mobility perturbations in cities during extreme weather events. *International Journal of Disaster Risk Reduction* **72**, 102849 (2022).

36. Abbou, A. *et al.* Household Adaptations to Infrastructure System Service Interruptions. *J. Infrastruct. Syst.* **28**, 04022036 (2022).
25

37. Davidson, R. A. *et al.* Typology of household adaptations to infrastructure system service interruptions. *International Journal of Disaster Risk Reduction* **97**, 103974 (2023).

38. Soleimani, N., Davidson, R. A., Kendra, J., Ewing, B. & Nozick, L. K. Household Adaptations to and Impacts from Electric Power and Water Outages in the Texas 2021 Winter Storm. *Nat. Hazards Rev.* **24**, 04023041 (2023).
26